# Winning the Big Data Technologies Horizon Prize: Fast and reliable forecasting of electricity grid traffic by identification of recurrent fluctuations

Jose M. G. Vilar[1,2,*]

[1] Biofisika Institute (CSIC, UPV/EHU), University of the Basque Country (UPV/EHU), P.O. Box 644, 48080 Bilbao, Spain

[2] IKERBASQUE, Basque Foundation for Science, 48011 Bilbao, Spain

[*]Correspondence: j.vilar@ikerbasque.org

## Abstract

This paper provides a description of the approach and methodology I used in winning the European Union Big Data Technologies Horizon Prize on data-driven prediction of electricity grid traffic. The methodology relies on identifying typical short-term recurrent fluctuations, which is subsequently refined through a regression-of-fluctuations approach. The key points and strategic considerations that led to selecting or discarding different methodological aspects are also discussed. The criteria include adaptability to changing conditions, reliability with outliers and missing data, robustness to noise, and efficiency in implementation.

## Introduction

There are currently highly significant problems in society, environment, industry, human health, and many other domains that would greatly benefit from the use of large amounts of historical and current data to predict the evolution of systems in real-time to act upon them to control their behavior [1]. In this type of situations, predictions need to be both accurate and fast to effectively implement the required actions.

Electricity grid traffic is a major example of such problems [2]. Electricity is not stored on a large scale. Therefore, matching supply with demand is critical to avoid wasting resources or risking power shortages. Doing so efficiently, however, is becoming increasingly challenging. Traditionally, variability was present only at the consumption end. Power plants could adjust their production to the anticipated consumer habits. With the widespread implementation of new renewable energy sources, such as wind turbines and solar plants, electricity generation has become also highly unpredictable. It depends on wind, sunlight, cloud coverage, and other weather conditions. The electric grid connects production with consumption and its traffic incorporates such unpredictable variability together with operational decisions and economic considerations. To match supply with demand, therefore, forecasting methods need to be able to make accurate predictions over the whole grid, including consumption, production, and operational effects.

The challenge of the Big Data Technologies Horizon Prize (http://ec.europa.eu/research/horizonprize/index.cfm?prize=bigdata) was to develop a software to predict over one-hour ahead the flow through a grid of 1912 high-tension lines at 5-minute intervals from large amounts of data with the possibility of considering a number of geospatial factors, such as weather and environmental parameters. The evaluation was based on the accuracy of the predictions, quantified by the root mean square error, and speed of the software, quantified by the overall elapsed execution time, with the ranking in accuracy given twice the weight of the ranking in speed to obtain the final score. The information available about the electricity flow was limited to just numerical values at specified times. The lines were anonymized with respect to location and there was no indication of a subjacent physical model that could be used nor how the lines were connected to each other.

To develop the software, the organization provided electricity flow data for a year of operations ( $\sim 2 \times 10^8$ data points) and specified the requirements that the software should meet. The data had missing values, outliers, and seem to have included linear interpolations over long periods of time for information. The actual evaluation of the software was made over undisclosed new sets of data. These conditions put strong constraints on the model assumptions since the software should autonomously train and adapt with such type of unseen data to make the predictions.

In general, electricity flow has high variability and their properties differ substantially from those of other quantities, such as electricity load or demand. The data related to the Big Data Technologies Horizon Prize contained restrictions that do not allow it to be reported herein. There are, however, a number of websites, as for instance http://www.elia.be/,

https://transparency.entsoe.eu, and https://www.bpa.gov, that have publicly posted electricity flow times series. Figure 1 shows, for illustrative purposes, a synthetically generated time series that emulates in just eight days some of the salient features of actual data: it can be positive or negative; it can have missing values; it can suddenly switch off to zero and back on from zero; it can have spikes; and it can be highly fluctuating, usually including daily and weekly components.

**Mathematical preliminaries**

A sensible strategy, based on the properties of the flow, is to consider the problem as short-term forecasting of continuous non-differentiable fluctuating coupled variables with multiple levels of seasonality. Short-term means explicitly one-hour-ahead predictions at 5-min intervals. Multiple levels of seasonality include daily and weekly components.

The simplest, trivial approach is to use the current value of the variable $y_t$ at time $t$ as a predictor $\hat{y}_{t+i}$ of the value of the variable a future time $t + i$:

$$\hat{y}_{t+i} = y_t \, .$$

This approach is referred to as the persistence model and assumes continuity of the variable. Since the variable is non-differentiable, it is not feasible to use the short-term trend, $(y_t - y_{t-1})i$, to improve the forecast. An alternative is to rely on seasonality to estimate the change in the variable as the change one day before, which leads to

$$\hat{y}_{t+i} = y_t + y_{t+i-\tau_D} - y_{t-\tau_D} \, ,$$

where $\tau_D$ is the time corresponding to one day. If seasonality involves also a weekly component, characterized by a time $\tau_W$, the prediction is expressed as

$$\hat{y}_{t+i} = y_t + (1 - g_t)(y_{t+i-\tau_D} - y_{t-\tau_D}) + g_t(y_{t+i-\tau_W} - y_{t-\tau_W}) \, ,$$

where $g_t$, with $0 \leq g_t \leq 1$, indicates the relative contributions of the daily and weekly components. Approaches based on seasonality have been shown to perform extremely well in forecasting country-wide global electric loads [3].

## Analysis of the data

To analyze the data provided for the development phase, the previous approaches were extended to consider short-term changes and changes multiple days before in the multivariate case of a grid. The prediction for the variables, denoted by $y_{k,t}$, is expressed as

$$\hat{y}_{k,t+i} = y_{k,t} + \sum_{l=1}^{L} b_{k,t,l}(y_{l,t} - y_{l,t-i}) + \sum_{l=1}^{L}\sum_{n=1}^{N} a_{k,t,l,n}(y_{l,t+i-n\tau_D} - y_{l,t-n\tau_D}).$$

Here, the coefficients $a_{k,t,l,n}$ refer to the effects of the changes of the $l$ variable $n$ days earlier on the $k$ variable at time $t$; $b_{k,t,l}$ considers the potential short-term dependence of the $k$ variable on the $l$ variable; the constant $N$ indicates the number of past days considered; and $L$ is the number of variables in the grid. Note that the univariate case is recovered when $a_{k,t,l,n} = a_{k,t,k,n}\delta_{k,l}$ and $b_{k,t,l} = b_{k,t,l}\delta_{k,l}$, where $\delta_{k,l}$ is the Kronecker delta function, which is 1 for $k = l$ and 0 for $k \neq l$.

If the coefficients $a_{k,t,l,n}$ and $b_{k,t,l}$ do not to depend on time, namely, if $a_{k,t,l,n} = a_{k,0,l,n}$ and $b_{k,t,l} = b_{k,0,l}$, it is possible to estimate their optimal values by performing regression analyses. The quantity to minimize is the mean square error,

$$MSE = \frac{1}{t_f - i - t_i - N\tau_D + 1} \sum_{t=t_i+N\tau_D}^{t_f-i} (y_{k,t+i} - \hat{y}_{k,t+i})^2,$$

for each variable $k$ subjected to different penalty functions and constraints in the coefficients $a_{k,0,l,n}$ and $b_{k,0,l}$. Here, $t_i$ and $t_f$ are, respectively, the initial and final time of the training set.

The data was split in training and test sets and the following cases were considered:

1. Univariate one-hour before value to predict one-hour change from the current value of every variable ($b_{k,0,l} = b_{k,0,k}\delta_{k,l}$, $N = 0$, and $i = 12$).
2. Multivariate one-hour before values of all the variables to predict one-hour ahead values from current values ($N = 0$ and $i = 12$).
3. Multivariate one-hour before values and one-hour changes one day before of all the variables to predict one-hour changes from current values ($N = 1$ and $i = 12$).
4. Univariate one-hour change one day before to predict one-hour change from the current value of every variable ($a_{k,0,l,n} = a_{k,0,k,n}\delta_{k,l}$, $b_{k,0,l} = 0$, $N = 1$, and $i = 12$).

5. Univariate one-hour changes multiple days before to predict one-hour change from the current value of every variable ($a_{k,0,l,n} = a_{k,0,k,n}\delta_{k,l}$, $b_{k,0,l} = 0$, $N = 7$, and $i = 12$).

Linear regression with univariate one-hour before values (case 1 above) was unable to outperform the persistence model on the test set. Taking one-hour before values of all the variables (case 2 above) performed substantially worse than the persistence model. It was a clear case of overfitting the noise that could not generalize. Indeed, using Lasso regression [4] reversed the sign of the outcome, slightly reducing the mean square error (MSE) of the persistence model in test set. This type of regularization led to sparse connections among the elements of the grid. Non-sparse methods, such as ridge regression [5], did not show such an improvement. Similarly, the results of restricting the couplings between variables to a neighborhood in a self-organizing map [6] systematically underperformed the persistence model in the absence of regularization.

Incorporation of one-hour changes one day before (case 3 above) in the Lasso regression approach substantially reduced the MSE of the persistence model in the test set. This increased performance results from taking into account recurrent daily changes. Intuitively, it means that changes in the grid traffic one day can be used to predict next-day changes at the same time.

The coefficients of the regression showed that the major predictors of the changes of one variable are the changes of the same variable one day before at the same time of the day. Indeed, considering only one-hour change one day before in a univariate approach (case 4 above) led only to slightly reduced performance with respect to the multivariate case. To test the limits of the single-variable approach, I considered together one-hour changes one day before, two days before, etc. (case 5 above). In this case, the performance matched that of the multivariate approach.

In addition to linear regression, gradient boosting regression [7] was also performed. The results were only marginally better than those of linear regression. One of the reasons for using gradient boosting was the possibility to incorporate environmental variables, such as temperature, wind speed, and radiation, in the regression approach. The incorporation of these variables, however, did not improve the performance.

## Main conclusion of the analyses

The analyses performed in the previous section indicate that a given variable typically exhibits recurrent patterns with random-like fluctuations and that fluctuations can be averaged out to some extent using other variables in the grid or the same variable at multiple one-day intervals in the past. The method finally used for the implementation of the software focused on single-variable prediction because, if sufficient past data is available, the potential advantage of using all the variables coupled together is likely to be offset by the complexity of lasso regression in a realistic situation.

## Forecasting through recurrent fluctuations

To perform forecasts based on averages of increments, it is useful to decompose the average increment in terms of single-step averages. These single-step averages can be viewed as the recurrent component of the fluctuations of the variable. With this decomposition, the prediction, $\hat{y}_{t+i|t}$, at time $t + i$ from time $t$ is performed as the sum of recurrent fluctuations using

$$\hat{y}_{t+i} = y_t + \sum_{j=1}^{i} \langle \Delta y_{t+j} \rangle,$$

with the expected recurrent fluctuation from time $t - 1$ to time t given by

$$\langle \Delta y_t \rangle = \frac{1}{N_t^*} \sum_{n=1}^{N} (y_{t-n\tau_D} - y_{t-1-n\tau_D}) I_{t-n\tau_D, t-1-n\tau_D}.$$

Here $I_{t,t'}$ is an indicator function for the validity of the $y_t - y_{t'}$ increment; namely, the indicator function is 1 if the increment consists of valid data and 0 otherwise. The normalization factor is formally defined as $N_t^* = \sum_{n=1}^{N} I_{t-n\tau_D, t-1-n\tau_D}$ and counts the number of valid increments used in the average.

Figure 2 shows the expected recurrent fluctuations, $\langle \Delta y_t \rangle$, and their one-hour (4-time-step) composition, $\sum_{j=1}^{4} \langle \Delta y_{t+j} \rangle$, for the last day of the time series shown in Figure 1 using the values of the 7 preceding days to compute the averages. These quantities have been computed for both the raw increments and the increments filtered to discard discontinuities, zeros, and spikes. Figure 3 shows the one-hour-ahead predictions with this approach compared with those of the persistence model and Figure 4 shows the same as figure 3 but for three-hour-ahead predictions. The results of using the composition of recurrent fluctuations clearly outperform those the persistence model despite the fluctuations, spikes, and switching off and back on the flow.

**Fast and reliable mathematical approach**

The strategy followed is based on using the recurrence of the fluctuations as a starting point to implement a robust, accurate, and efficient method that could adapt to changing conditions and process the incidents of real data.

The first modification of the basic approach is to use exponential smoothing as an adaptive efficient method for averaging. With this method, the expected value $D_t$ of the daily recurrent fluctuation of the variable $y_t$ from time $t-1$ to time $t$ is computed as

$$D_t = r_D I_{t,t-1}(y_t - y_{t-1}) + (1 - r_D I_{t,t-1}) D_{t-\tau_D},$$

where $\tau_D = 288$ is the number of 5-minute time increments in one day and $r_D$ determines the persistence of the moving average. Notice that $I_{t,t'}$ is the indicator function for the validity of the $y_t - y_{t'}$ increment as defined in the previous section. If a problematic value of the increment is encountered at time $t$, $I_{t,t'} = 0$ and consequently $D_t = D_{t-\tau_D}$, leaving $D_t$ unchanged. The values of $D_t$, as well as those of the other moving averages below, are always well-defined.

The second improvement is to consider also weekly recurrent fluctuations, computed through

$$W_t = r_W I_{t,t-1}(y_t - y_{t-1}) + (1 - r_W I_{t,t-1}) W_{t-\tau_W},$$

to potentially increase the prediction accuracy of variables that also depend on the day of the week. Here $\tau_W = 288 \times 7$ is the number of 5-minute time increments in one week and $r_W$ determines the persistence of the moving average.

With these two quantities, a preliminary prediction, $\hat{y}^{(1)}_{t+i|t}$, at time $t+i$ from time $t$ is performed as

$$\hat{y}^{(1)}_{t+i|t} = y_t + (1 - g_t)\sum_{j=1}^{i} D_{t-\tau_D+j} + g_t \sum_{j=1}^{i} W_{t-\tau_W+j},$$

where the coefficient $g_t$ weights the relative contribution of each type of recurrent fluctuation and is computed dynamically to minimize the moving square error of past predictions at 11 time steps (55 min). The explicitly equations used are

$$g_t = \frac{g_t^C}{g_t^V},$$

$$g_t^C = r_g I_{t,t-11}(y_t - y_{t-11} - sD_{11})(sW_{11} - sD_{11}) + (1 - r_g I_{t,t-11})g_{t-1}^C,$$

$$g_t^V = r_g I_{t,t-11}(sW_{11} - sD_{11})^2 + (1 - r_g I_{t,t-11})g_{t-1}^V,$$

with $sD_i = \sum_{j=1}^{i} D_{t-i-\tau_D+j}$ and $sW_i = \sum_{j=1}^{i} W_{t-i-\tau_W+j}$. Here, $r_g$ determines the persistence of the moving averages. In mathematical terms, these equations provide an efficient adaptive way to compute the value of $g_t$ that minimizes $\sum_{j=11}^{t} r_g^{t-j}(y_j - \hat{y}^{(1)}_{j|j-11})^2$.

The approach also monitors the moving square error of past predictions at 11 time steps with a characteristic persistence parameter $r_e$ and that of the corresponding predictions of the persistence model:

$$e_t^M = r_e I_{t,t-11}(y_t - \hat{y}^{(1)}_{t|t-11})^2 + (1 - r_e I_{t,t-11})e_{t-1}^M,$$

$$e_t^P = r_e I_{t,t-11}(y_t - y_{t-11})^2 + (1 - r_e I_{t,t-11})e_{t-1}^P.$$

At a given time, a second preliminary prediction, $\hat{y}^{(2)}_{t+i|t}$, is made as

$$\hat{y}^{(2)}_{t+i|t} = \hat{y}^{(1)}_{t+i|t} \text{ if } e_t^M \leq e_t^P,$$

$$\hat{y}^{(2)}_{t+i|t} = y_t \text{ if } e_t^M > e_t^P.$$

This ensures that the approach will not be making predictions worse than those of the persistence model for a long time.

The approach also incorporates a correction to the second preliminary prediction. It is a term proportional to the difference between the current value of $y_t$ and that of its moving average, $m_t$. It is given by $o_t(m_t - y_t)i$ and is analogous to the regression of fluctuations in statistical physics. The value of $o_t$ is computed by minimizing the moving residual square error of past predictions. Explicitly, a third preliminary prediction is made as

$$\hat{y}^{(3)}_{t+i|t} = \hat{y}^{(2)}_{t+i|t} + o_t(m_t - y_t)i,$$

with

$$m_t = r_m I_t y_t + (1 - r_m I_t) m_t,$$

$$o_t = \frac{o_t^C}{o_t^V},$$

$$o_t^C = r_o I_{t,t-11} \frac{1}{11}(y_t - y_{t-11} - (1 - g_t)sD_{11} - g_t sW_{11})(m_t - y_t) + (1 - r_o I_{t,t-11})o_{t-1}^C,$$

$$o_t^V = r_o I_{t,t-11}(m_t - y_t)^2 + (1 - r_o I_{t,t-11})o_{t-1}^V.$$

Here, as usual, $r_m$ and $r_o$ determine the persistence of the moving averages and $I_t$ is the indicator function for the validity of the value of $y_t$. It is important to note that $I_{t,t'}$, the indicator function for the validity of the $y_t - y_{t'}$ increment, is more stringent than $I_t$.

The final prediction is made by damping high changes through a saturation function as

$$\hat{y}_{t+i|t} = y_t + \frac{\hat{y}^{(3)}_{t+i|t} - y_t}{1 + \frac{1}{K_s}\left|\hat{y}^{(3)}_{t+i|t} - y_t\right|},$$

were $K_s$ is the maximum change allowed.

## Software implementation

The software was implemented in Python, using NumPy for vector and matrix operations. Because of efficiency purposes, only the values of the variables needed in the equations were stored. This implies one value for variables that are updated from the previous time step, 288 values from variables that are updated from the previous day, and $288 \times 7$ values from variables that are updated from the previous week.

The moving averages are computed by discarding problematic values. Namely, the variable accounting for the average is not updated at the problematic time step, as described in the equations above. Problematic values include changes of variables over a threshold bigger than two times the root mean square value of all the variables, changes of variables with gaps in time, non-available values (NaNs), outliers, variables that are zero for at least two consecutive time points, and missing time points.

The software also ensured that the terminal value $y_t$ used to make the prediction $\hat{y}_{t+i|t}$ was well defined. If it is missing or an outlier, it was computed by extrapolation of previous well-defined values. If no well-defined value is available, it is set to zero. The software was programmed to default to the persistence-model prediction and to try to recover by reinitiating the moving averages if any unexpected problem occurred during the computations. Time was always converted to 5 min time steps. If a time point would fall between two time-steps, it will be truncated and assigned to the lowest value.

The reliability and accuracy of the software were tested with the original data provided by the organization and also by simulating new scenarios through targeted modifications of the datasets provided. The modifications included the addition of multiple types of outliers, removal of a large number of time points, shuffling of data, and changing the number of variables through the training, adaptation, and testing processes.

## Discussion

The Big Data Technologies Horizon Prize has provided an unprecedented opportunity to assess and expand our limits for developing reliable software that autonomously trains and adapts on large amounts of actual data to accurately predict electricity grid traffic. These types of data-driven problems are highly significant in many domains, including society, environment, industry, and human health.

The challenges involved are two-fold. On the one hand, there is the software engineering side. The end product should be resilient to the incidents of actual data, which cannot realistically be preprocessed by humans in most real-time scenarios. In addition, it should deliver the results sufficiently fast. On the other hand, there is the mathematical side. Methodologies should be designed to enable such resiliency and speed at the same time as being as accurate as possible.

To tackle these challenges, the strategy I followed consisted in using the state of the art of current methodologies to analyze the data and to identify key determinants of the predictability of the system behavior. Subsequently, I developed the simplest possible methodology that could perform on the available data at the same level as the most accurate approaches. Discarding complexity that leads only to marginally better accuracy on the available data is typically beneficial for efficiency and, most importantly, for reliability [8]. Unless correctly managed, complexity has proved to be one of the leading causes of overfitting that prevents generalization. Simulating new scenarios through targeted modifications of the available datasets was invaluable to discard potentially problematic methodological aspects and to enable the software to reliably perform on conditions not present in the original datasets.

The selected approach relied on identifying the expected fluctuation of the flow at a given time from fluctuations at the same time of the day on earlier days and how this fluctuation is affected by the departure of flow from its average value.

The initial analysis on the available data indicated that considering the recent past of the whole grid to make predictions on a given line was only marginally more accurate than focusing on longer historical data of the line. A potential avenue for improvement would be to incorporate the structure of the grid into a predictive model that could consider sufficient historical data of multiple lines to make predictions on a given line. The major challenge of this approach was not to build the model itself but to do so autonomously in an adaptive manner on undisclosed data with potentially unknown and changing properties.

The initial analysis also indicated that the incorporation of environmental variables, such as temperature, wind speed, and radiation, did not improve the performance, which is consistent with the possibility that the information provided by these variables is already encoded in the grid data. A potential avenue for improvement in this direction would be to consider reliable forecasts of the environmental variables to incorporate information about the future that is not present in the grid.

## Acknowledgments

I would like to thank the European Commission Directorate-General for Communications Networks, Content and Technology for supporting the initiative that led to this work and all the persons involved in the organization of the Big Data Technologies Horizon Prize. This participant was one of the winners of the Inducement Prize: Big Data technologies from the European Union's Horizon 2020 research and innovation programme.

# Figures

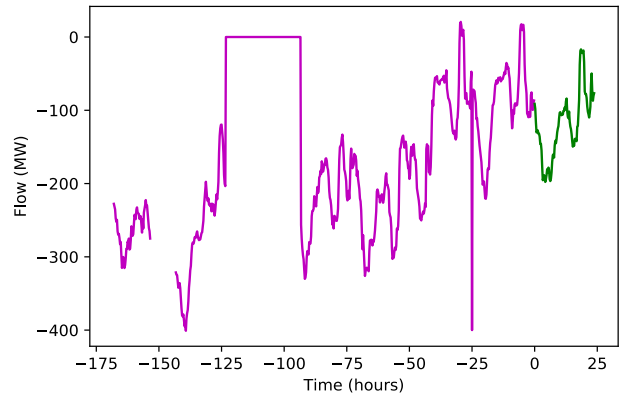

**Figure 1:** Synthetically generated time series that emulates electricity flow data. The time step is 15 min. The last day, starting at time 0, is displayed in a different color than the previous seven days.

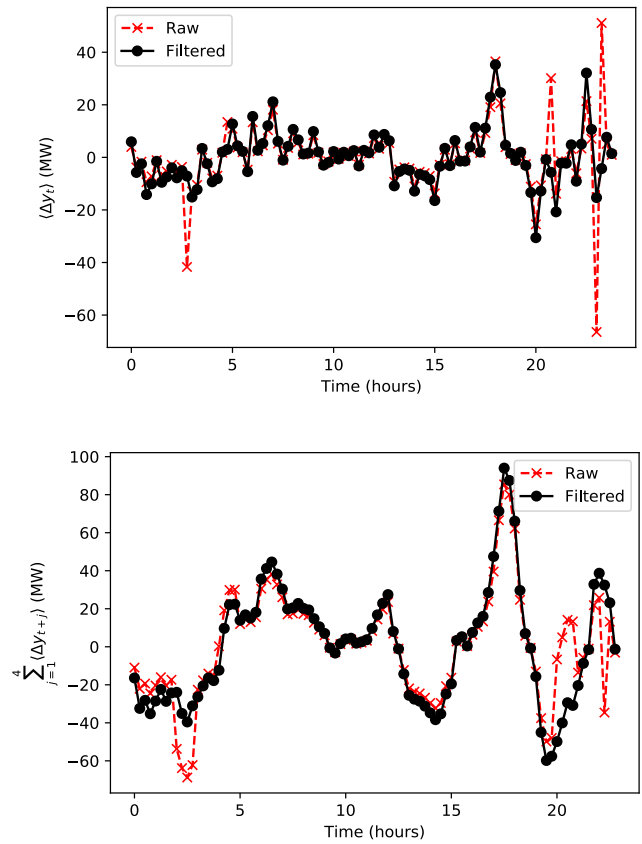

**Figure 2:** Expected recurrent fluctuations (top) and their one-hour composition (bottom) for the last day in Figure 1 as functions of time for both the raw increments (x symbols) and the increments filtered to discard discontinuities, zeros, and spikes (filled circles). The time step is 15 min.

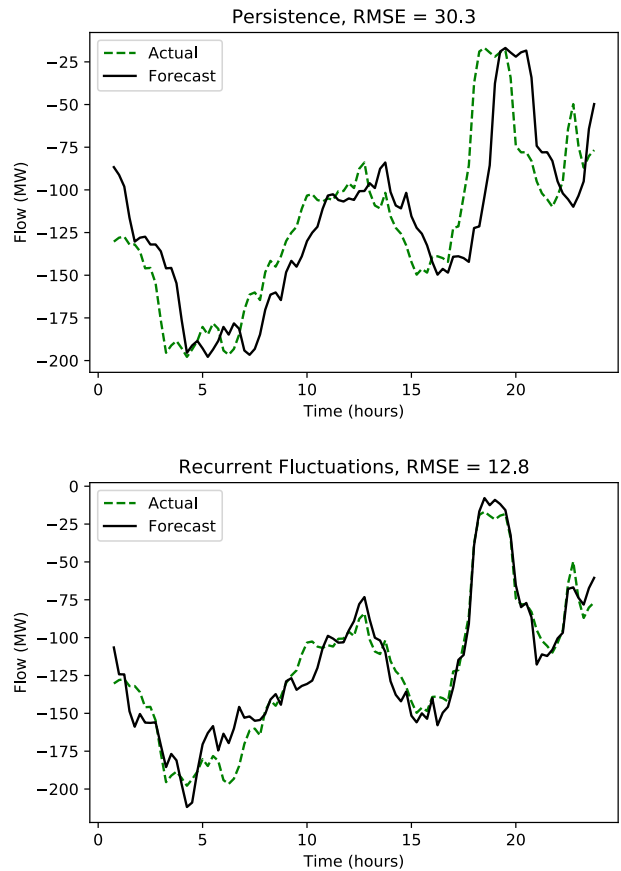

**Figure 3**: One-hour-ahead predictions (continuous lines) for the last day in Figure 1 with the persistence model (top) and with the recurrent fluctuations approach (bottom) compared to the actual data (dashed lines). The error of the prediction is indicated on the top of each panel.

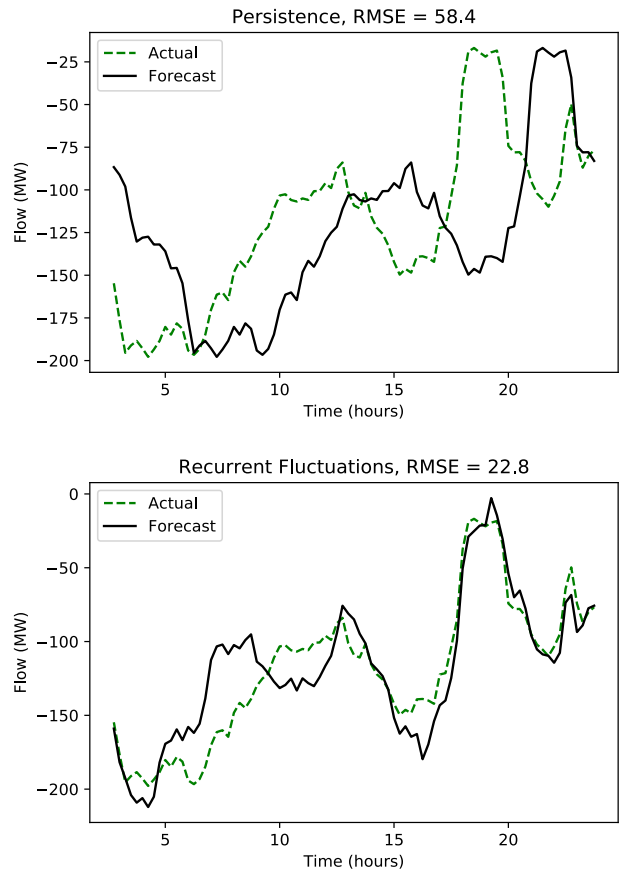

**Figure 4**: Three-hour-ahead predictions (continuous lines) for the last day in Figure 1 with the persistence model (top) and with the recurrent fluctuations approach (bottom) compared to the actual data (dashed lines). The error of the prediction is indicated on the top of each panel.